\documentclass[fleqn,10pt]{wlscirep}

\usepackage{csquotes}
\usepackage{bm}
\usepackage{scrextend}
\usepackage{multirow}
\usepackage{xr}
\newcommand\defeq{\mathrel{\overset{\makebox[0pt]{\mbox{\normalfont\tiny\sffamily def}}}{=}}}

\title{A Preferential Attachment Paradox: How Preferential Attachment Combines with Growth to Produce Networks with Log-normal In-degree Distributions}

\author[1,*]{Paul Sheridan}
\author[2]{Taku Onodera}
\affil[1]{Hirosaki University, Department of Active Life Promotion Science, Hirosaki, 036-8562, Japan}
\affil[2]{The University of Tokyo, Institute of Medical Science, Human Genome Center, Tokyo, 108-8639, Japan}

\affil[*]{paul.sheridan.stats@gmail.com}

\begin{abstract}
Every network scientist knows that preferential attachment combines with growth to produce networks with power-law in-degree distributions. How, then, is it possible for the network of American Physical Society journal collection citations to enjoy a log-normal citation distribution when it was found to have grown in accordance with preferential attachment? This anomalous result, which we exalt as the preferential attachment paradox, has remained unexplained since the physicist Sidney Redner first made light of it over a decade ago. Here we propose a resolution. The chief source of the mischief, we contend, lies in Redner having relied on a measurement procedure bereft of the accuracy required to distinguish preferential attachment from another form of attachment that is consistent with a log-normal in-degree distribution. There was a high-accuracy measurement procedure in use at the time, but it would have have been difficult to use it to shed light on the paradox, due to the presence of a systematic error inducing design flaw. In recent years the design flaw had been recognised and corrected. We show that the bringing of the newly corrected measurement procedure to bear on the data leads to a resolution of the paradox. 
\end{abstract}
\begin{document}

\flushbottom
\maketitle
%
%
\thispagestyle{empty}


\section*{Introduction\label{SEC_Introduction}}

The physicist Sidney Redner reported a rather curious anomaly in a decade-old study on the citation statistics of the \emph{American Physical Society} (APS) journal collection~\cite{Redner2005}. Redner effectively discovered that while the APS citation network had grown in accordance with a process commonly know as preferential attachment, the corresponding citation distribution closely follows a log-normal distribution on a double logarithmic scale. The network scientist will recognise preferential attachment as a process whereby the nodes of a network acquire new connections in proportion to the number of connections they already entertain. What makes his observations so puzzling is that growing network models based on preferential attachment have long been known to generate networks with power-law, as opposed to log-normal, in-degree distributions~\cite{deSollaPrice1976, Barabasi1999, Dorogovtsev2000, Krapivsky2000, Krapivsky2001}. This anomaly, or paradox, as Redner referred to it, may be called for convenience the \emph{preferential attachment paradox}.

In this paper we propose a resolution to the paradox. But we first take pains to reproduce the anomalous findings that Redner reported.~In so doing we confirm that the APS citation distribution closely follows a log-normal distribution on a double logarithmic scale, and, moreover, that the associated APS citation network had grown in accordance with preferential attachment. Only then do we venture to resolve the paradox. The resolution we propose requires that two main obstacles be overcome. The first is to recognise that whether preferential attachment is observed in a growing network or not depends on the choice of measurement procedure. This insight will lead us to conclude that the preferential attachment observed by Redner amounts to an artefact of the procedure he used to measure the process over coarse time resolutions~\cite{Jeong2003}. And, what is more, when we perform the measurements at comparatively fine time resolutions, the outcomes are found to be reconcilable with a form of attachment that is consistent with a log-normal citation distribution. The second obstacle is a purely technical matter related to the accurate measurement of preferential attachment at fine time resolutions. The fact that the paradox has remained unresolved until now is explained in part by the presence of a design flaw in the standard fine time resolution measurement procedure~\cite{Newman2001}. The flaw, for which a correction has only recently been suggested~\cite{Thong2015}, has been an obstacle to progress, because it functions to distort the measurements taken using the procedure, much as a crooked ruler distorts the true lengths of the objects that it measures. But once these obstacles to measuring preferential attachment have been realised, we will see how a little detective work is all that stands in the way of a definitive resolution to the paradox.

\section*{The Preferential Attachment Paradox\label{SEC_PA_Paradox}}

In the previous section we outlined the preferential attachment paradox. Informally it is this:
\begin{displayquote}
Growing network models based on growth and preferential attachment are known to generate networks with power-law in-degree distributions. So how can preferential attachment combine with growth to generate networks with log-normally distributed in-degree distributions? In particular, how did preferential attachment give rise to a log-normally distributed citation distribution in the growth of the APS citation network?
\end{displayquote}
In this section we formulate the paradox in technical terms and illustrate it on the APS journal collection citation network. The illustration is in point of fact a careful reproduction of those anomalous results reported by Redner in the very same paper in which he first called attention to the paradox~\cite{Redner2005}. For the sake of review they are: 1) the APS citation distribution closely follows a log-normal distribution on a double logarithmic scale, and 2) the associated APS citation network had grown in accordance with preferential attachment. We reproduce these results successively below. But we begin with an overview of the APS journal collection citation data.

\subsection*{The APS Journal Collection Citation Data\label{SUBSEC_APS_Data}}

The APS ranks among the world's foremost learned societies for physicists. The society publishes a dozen research journals that span virtually all fields of modern physics. Its journal collection citation data from July 1893 through December 2009 is freely available for download upon request at the society website~\cite{APS_Data2016}.

The dataset is comprised of just over $450,000$ timestamped articles and $4,500,000$ intra-APS journal citations. But in keeping with Redner, whose analysis we aim to reproduce, we restrict our attention to only those articles from July 1893 up to and including June 2003. According to our tally this $110$ year stretch of data covers precisely $347,038$ articles and $3,063,726$ citations. The mean number of citations is $8.8$ which agrees with Redner's reported value. The scrupulous reader may object that Redner reports $353,268$ articles and $3,110,839$ citations over the same time period. In other words, there is about a $1\%$ shortfall on our part in both instances. Roughly $40\%$ of the missing citations are accounted for by the fact that we filtered $12,425$ duplicate citations and $115$ self-citations in the course of processing the data. The remaining shortfalls are perhaps attributable to vigorous data cleansing efforts on the part of APS technicians over the years.

Any bibliographic dataset is readily conceptualised as a type of network called a citation network. The nodes of a citation network represent articles in such a manner that a node Y is connected to a node X if the article corresponding to X is cited by the article corresponding to Y in its references. The said connection, if it exists, is conferred with an orientation so as to point like an arrow from the citing article Y to the cited article X. Multiple connections from Y to X (i.e. duplicate citations) and self-loops from Y to Y (i.e. self-citations) are prohibited as a matter of convenience. Thus a citation network, at least in this paper, will be recognised by network aficionados as a simple directed network representation of bibliographic data. And when we speak of the APS citation network without qualification, we mean precisely this kind of representation of the APS citation data from July 1893 through June 2003 as related above.

\subsection*{The APS Citation Network Citation~Distribution\label{SUBSEC_APS_Citation_Distribution}}

The distribution of node degrees in a network is one of the most important network properties, and a defining characteristic of network structure. Many readers will already be familiar with the notion that the \emph{in-degree} $k$ of a node in a directed network is the number of incoming connections it shares with other nodes, and moreover that the \emph{in-degree distribution}, $P(k)$, is an associated function which gives the proportion of nodes in the network with in-degree~$k$. In the context of citation networks, a usual goal of the network scientists is to characterise the distribution of incoming citations. They give the name \emph{citation distribution} to the in-degree distribution $P(k)$ of a citation network, which is at once seen to give the proportion of papers in the network cited $k$ times. Network scientists have long explored fitting citation distributions by a variety of different functional forms; see Radicchi et al.~\cite{Radicchi2012} for a brief review. Suffice it to say here that the question of which functional form\thinspace--\thinspace if any\thinspace--\thinspace best characterises citation distributions remains a subject of ongoing research.

Redner appealed to the log-normal to describe the APS citation distribution in his study of citation statistics from the first 110 years of the APS journal collection~\cite{Redner2005}. Strictly speaking, he found that visual inspection reveals the (\emph{complementary}) \emph{cumulative in-degree distribution} $C(k) = \sum_{i \geq k} P(i)$ of the APS citation distribution $P_{\mbox{\tiny APS}}(k)$ is well-fitted by the so-called log-normal form $\mathcal{L}(k;\beta_0,\beta_1,\beta_2) = \beta_0 \exp \left[ -\beta_1 \log(k) -\beta_2 \log^2(k) \right]$ over a substantial range of incoming citations when $\beta_0 = 0.15$, $\beta_1 = 0.40$, and $\beta_2 = 0.16$. In the context of a citation network, we refer to $C(k)$, which gives the proportion of papers cited at least $k$ times, as a \emph{cumulative citation distribution}. Redner concluded on the basis of the above outcome that the form of $P_{\mbox{\tiny APS}}(k)$ is inconsistent with a power-law for reasons described in Supplementary Note~1. 

The remainder of this section is devoted to showing that $P_{\mbox{\tiny APS}}(k)$ is better described by a discretisation of the log-normal distribution
\begin{equation}~\label{EQ_Lognormal_Pk}
\log \mathcal{N}(k; \mu, \sigma) =  \frac{1}{(k+1) \sqrt{2 \pi \sigma^2}} \times \exp \left[ -\frac{\left(\log(k+1) - \mu \right)^2}{2\sigma^2} \right]
\end{equation}
for $k \geq 0$ with location parameter $\mu$ and scale parameter $\sigma > 0$, than by either of a corresponding discretised power-law $(k+1)^{-\gamma}$ or discretised exponential distribution $\lambda e^{-\lambda (k+1)}$ with rate parameter $\lambda > 0$. In addition, we investigate the extent to which the log-normal can be said to plausibly model $P_{\mbox{\tiny APS}}(k)$ in absolute terms. We use $k+1$ instead of $k$ in Eq.~(\ref{EQ_Lognormal_Pk}) on the one hand so as to include papers with zero citations, and on the other because doing so dovetails with the modelling framework that we will develop in a later section. The exponential distribution we include in our analysis as a token light-tailed alternative to the heavy-tailed log-normal and power-law distributions.

\begin{figure}[!h]
\centering
\includegraphics[width=8cm]{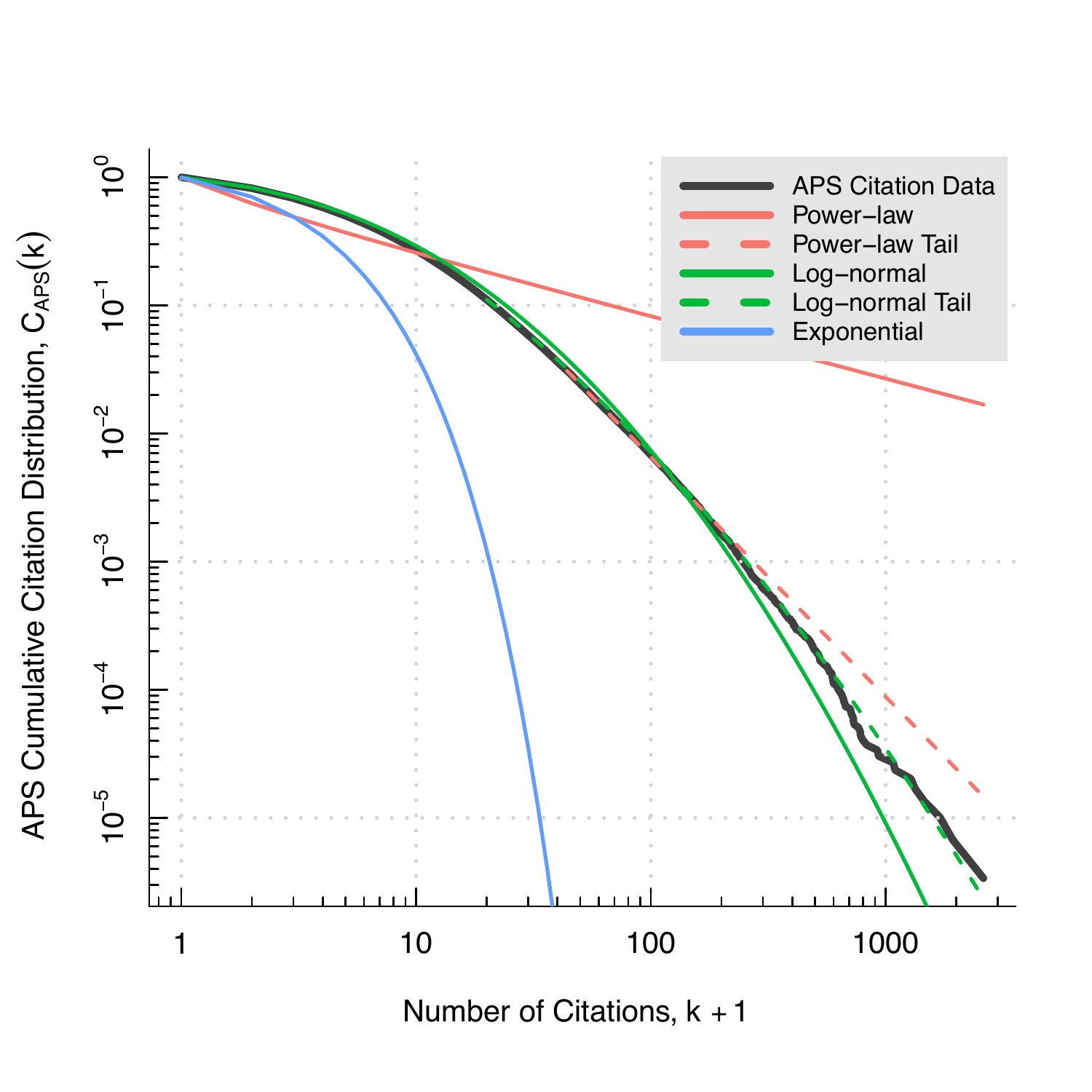}
\caption{\textbf{The APS cumulative citation distribution}~$C(k)$ for all publications dating from July 1893 up to and including June 2003. The log-normal cumulative distribution (green) fits the observed data better than either a power-law cumulative (red) or exponential cumulative distribution (blue). This holds true regardless of whether the data is fit over its full domain (solid lines) or merely in the tail region of $k$ (dashed lines). The cutoff values defining the tail regions (dashed lines) are calculated using the maximum likelihood estimation method of Clauset et al.~\cite{Clauset2009} described in the main text. \label{FIG_APS_Citation_Distribution}}
\end{figure}

We used the poweRlaw R package v.\thinspace0.60.3~\cite{Gillespie2015} to fit our trio of functional forms to the APS citation distribution. The package implements the maximum likelihood estimation methods and goodness-of-fit tests, based on the Kolmogorov-Smirnov (KS) test and likelihood ratios, for fitting heavy-tailed distributions to observed data described in Clauset et al.~\cite{Clauset2009}. Their approach may be summed up as follows: Candidate functional forms are separately fitted to an observed in-degree distribution over the domain $k + 1 \geq k_{min}$ by maximum likelihood, conditional on the choice of lower cutoff $k_{min}$. An optimal $k_{min}$ is estimated for each candidate using a goodness-of-fit based testing approach. Alternately, the value of $k_{min}$ may be set to $1$ to fit the data over its full domain. The goodness-of-fit of each functional form is assessed using a KS based hypothesis testing approach.

The APS cumulative citation distribution is plotted in Fig.~\ref{FIG_APS_Citation_Distribution} on a double logarithmic scale together with a medley of fitted functional form cumulatives. The associated cumulative distributions are plotted in Fig.~\ref{FIG_APS_Citation_Distribution} solely on aesthetic grounds. That said, a cursory visual inspection will satisfy even the most quantitatively minded reader that the log-normal better describes the entire APS citation distribution ($k_{min}$=1), than either a power-law or exponential distribution. And just like that our modest claim is proved correct.

This is by no means to say that the log-normal plausibly describes the APS citation distribution. On the contrary, we found that the goodness-of-fit of log-normal to the data ($H_{0}$ = log-normal with $\mu$ = 1.41, $\sigma$ = 1.27, $k_{min}$ = 1: KS = 0.01 \& P = 0.00) was poor insofar as rigorous statistical testing is concerned. Clauset et al.~\cite{Clauset2009} recommend a significance level of $0.10$ as a conservative choice for pronouncing a given functional form a plausible fit to the data\thinspace---\thinspace for what it is worth. Thus the hypothesis that the log-normal plausibly fits the APS citation distribution over its full domain is to be rejected.

But proponents of the log-normal will be heartened to learn that this unhappy state of affairs is entirely reversed once we confine our attention to either the body or tail region of the distribution. Let us take the APS citation distribution body and tail to correspond to the region $0 \leq k \leq 150$ and $k \geq 20$, respectively. Visual inspection of the Fig.~\ref{FIG_APS_Citation_Distribution} plot alone is enough to conclude that the body is plausibly fit by the log-normal distribution. We visually selected $k \leq 150$ as a conservative choice of cutoff point on account that the poweRlaw R package cannot be used to assess the goodness-of-fit of a log-normal to the body of a distribution. On the other hand, we found that the APS citation distribution tail is plausibly fit by the log-normal distribution at significance level 0.10 ($H_{0}$ = log-normal with $\mu$ = -1.00, $\sigma$ = 1.76, $k_{min}$ = 20: KS = 0.00 \& P = 0.33). Note the cutoff $k_{min} = 20$ is the minimum $k$ yielding a plausible fit of the log-normal to the data at significance level $0.10$. For this reason, we use $k \geq 20$ to define the tail of the distribution. The same, however, cannot be said for a power-law ($H_{0}$ = power-law with $\gamma$ = 2.87, $k_{min}$ = 44: KS = 0.01 \& P = 0.01). The plot of Fig.~\ref{FIG_APS_Citation_Distribution} serves to visually reinforce these conclusion. 

Let us conclude by taking stock of our findings. The log-normal distribution, we found, provides an incontrovertibly better fit to the APS citation distribution over its full domain, than does a power-law. Moreover, the log-normal looks to fit the APS citation distribution pretty nearly as judged by visual inspection, but this is not supported by rigorous statistical testing. This is because the log-normal undershoots the target in the tail of the distribution. We may nevertheless speak informally of the APS citation distribution as ``closely'' following a log-normal in certain non-technical contexts. That said, we found the log-normal does provide a good fit (in the technical sense) to the data when confined to either the body ($0 \leq k \leq 150$) or tail ($k \geq 20$) of the distribution. We are careful to be precise about which region of the distribution that we mean in technical contexts. In light of these considerations, the reader will do well to keep the informal and technical senses of the log-normal providing a close fit to the APS citation distribution in mind.

\subsection*{The APS Citation Network Attachment Rate\label{SUBSEC_APS_Attachment_Rate}}

A growing network represents bibliographic data over time in a manner conducive to the quantification of preferential attachment. Before considering how to represent bibliographic data as a growing network it should be understood that what we mean by bibliographic data is a collection of intra-referencing articles complete with timestamps of the form YYYY-MM-DD. While timestamps proved superfluous to the construction of the APS citation network, to a growing network representation of the APS bibliographic data they are essential. This is because a \emph{growing network} is formally defined as a nested sequence of networks, $\mathcal{G} = \left\{ G_t \right\}_{t=1}^T$, that begins with an \emph{initial network}, $G_1$, with $n_1 > 0$ nodes and $m_{1}' \geq 0$ edges and ends with a \emph{final network}, $G_T = G$. Nesting means that the network $G_t$ at \emph{time-step} $t$ for $t > 1$ is obtained by augmenting $G_{t-1}$ with $n_t \geq 0$ nodes that form $m_t \geq 0$ connections with the nodes in $G_{t-1}$ and $m_{t}' \geq 0$ connections among the nodes newly added (see Figure~S1 for a graphical depiction of this modelling scheme). A bibliographic dataset is represented as a growing network by specifying a mapping from article timestamps to sequence time-steps that preserves chronological order up to a desired level of time-resolution. Articles are mapped to nodes and references to directed edges within this framework in the obvious fashion. 

A few examples of growing network representations will serve to make their workings more comprehensible.~Table~S1 summarises the examples here described. The APS citation network is a growing network in the trivial sense that all article timestamps from 1893-07-01 to 2003-06-30 are mapped to a single time-step. The network in this case consists of $n_1 = 347,083$ nodes and $m_{1}' = 3, 063, 726$ edges. It is sometimes convenient to qualify the APS citation network as being \emph{minimally resolved} to emphasise its growing network nature. What may be called the \emph{maximally resolved} APS citation network falls at the opposite end of the time resolution spectrum. In this case there are as many time-steps as there are articles so that the sequence is grown by $n_t = 1$ node with $m_t \geq 0$ edges at each time-step $t$. The value of $m_{t}'$ is equal to $0$ for all~$t$ since self-citations are prohibited. Identically timestamped articles are discriminated according to the lexicographical ordering of their unique article IDs in a slight abuse of the representation. It is easy to imagine in a similar vein \emph{daily}, \emph{monthly}, and \emph{yearly resolved} APS citation networks lying between these two extremes. For example, yearly resolution means that all articles published in the same calendar year are mapped to nodes in the same time-step. For a given time-step $t$, $n_t$ is the number of articles published in the corresponding year, $m_t$ the number of citations to articles from previous years, and $m_t'$ the number of citations to articles in the same year. Note that journal issue print date timestamps are used to construct the APS citation network at daily resolution. Figure~S2 shows a conceptual depiction of the time resolutions here described. The growing networks we have described here will prove key to resolving the preferential attachment paradox in a later section.

\begin{figure}[!h]
\centering
\includegraphics[width=8cm]{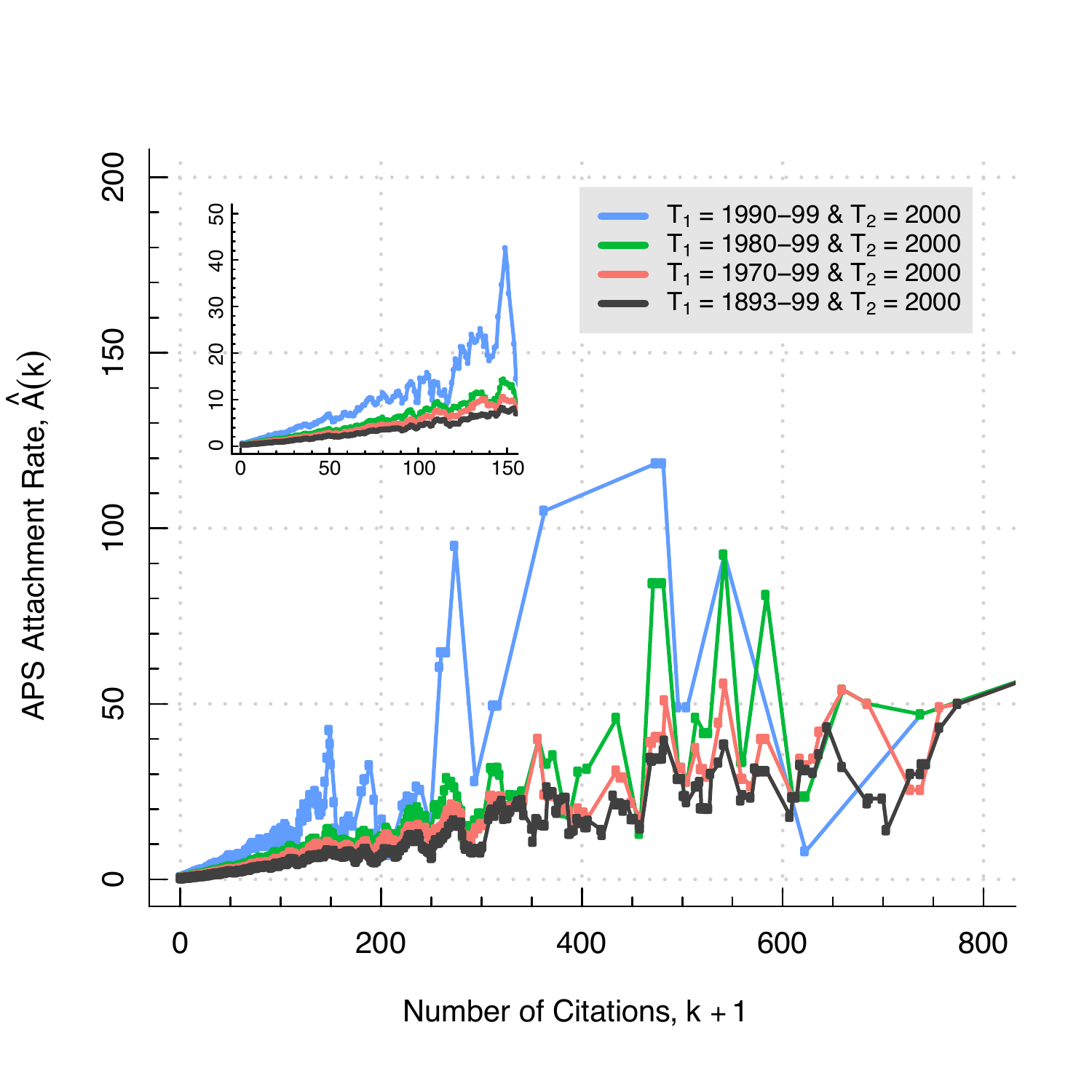}
\caption{\textbf{The attachment rate for various bi-epochally resolved APS citation networks} appear linear to the naked eye over a wide range of~$k$, especially in the region from $k=0$ to about $150$ (inset). Four different measurements of $\hat{A}(k)$ may be distinguished by colour in the plot. Each individual measurement is determined by recording how the publications in time interval $T_2=2000$ cite the publications in time intervals $T_1=1990$\thinspace-\thinspace$99$ (blue), $T_1=1980$\thinspace-\thinspace$99$ (green), $T_1=1970$\thinspace-\thinspace$99$ (red), and $T_1=1893$\thinspace-\thinspace$99$ (black), respectively. The data have been averaged over a range of $k \pm 0.025 k$. See Table~S2 and the surrounding text for~details.\label{FIG_APS_Attachment_Rate}}
\end{figure}

But in the present subsection, our focus is squarely on a collection of \emph{bi-epochally resolved} APS citation networks. In principle, bi-epochal resolution describes the scenario when a partition of the article timestamps of a bibliographic dataset into two non-overlapping intervals, labeled $T_1$ and $T_2$ hereafter, is used to define a growing network representation comprised of two time-steps. In practice, the time intervals do not always cover the entire data. Table~S2 summarises our reconstruction of four bi-epochally resolved growing network representations of the APS citation data that Redner submitted to analysis in his original study~\cite{Redner2005}.

In order to characterise preferential attachment network scientists measure the rates at which articles with $k$ citations are cited by new articles. This they achieve by observing the process of citation formation over time as viewed through the prism of this or that growing network representation. Loosely speaking, the \emph{attachment rate} $\hat{A}(k)$ of a growing network is defined as the likelihood that an edge from among the $m_t$ added at time-step $t>1$ connects to a node of in-degree~$k$. Jeong et al.~\cite{Jeong2003} proposed to measure the attachment rate of a bi-epochally resolved growing network, $\mathcal{G} = \{ G_1 , G_2 \}$, as 
\begin{equation}~\label{EQ_AR_Jeong}
\hat{A}(k) \defeq \frac{n_1}{m_2} \times \frac{m_{2}(k)}{n_{1}(k)},
\end{equation}
where $m_{2}(k)$ is the number of edges from $G_2$ that connect to an in-degree $k$ node in $G_1$ of which there are assumed to be $n_{1}(k)$ in number. The factor $n_1/ m_2$ serves as a mathematically convenient constant of normalisation. The domain is taken to be $\{k \mid m_{2}(k) / n_{1}(k) \neq 0 \}$ under the convention that $0/0=0$. Mark Newman took steps to generalise Jeong's measure to certain arbitrarily time resolved growing networks~\cite{Newman2001}. The details of Newman's measure are deferred to a later section. The attachment rate of a growing network is sometimes said to be ``preferential'' when the trend line of the measured $\hat{A}(k)$ is found to be an increasing function of $k$. But in this paper, we apply the term ``preferential'' to measured attachment rates in a more restricted sense. Namely: if $\hat{A}(k)$ increases linearly in $k$, then attachment rate is said to be \emph{preferential}. In this idealised case the in-degree distribution of the resulting network is bound to follow a power-law under certain regularity conditions~\cite{deSollaPrice1976, Barabasi1999, Krapivsky2000}.

Redner found the attachment rates for bi-epochally resolved growing network representations of various APS citation data subsets to be nearly linear functions of $k$ with the agreement being especially pronounced for $k$ less than $150$~\cite{Redner2005}. In Fig.~\ref{FIG_APS_Attachment_Rate} we reproduce his experimental measurements in all but a few extraneous details that are discussed in Supplementary Note~2. The attachment rate $\hat{A}(k)$, as defined by Jeong's measure, is plotted for bi-epochally resolved growing network representations of four different subsets of the APS citation data. In each case, the measured $\hat{A}(k)$ is observed to approximately follow a straight line. Thus, the presence of preferential attachment is apparently confirmed in each growing network. Redner extrapolates from these bi-epochal outcomes that preferential attachment accounts for the formation of citations in the maximally resolved APS citation network. Let us provisionally accept this conclusion with the understanding that it will be overturned in due course.

\subsection*{A Digression on Growing Network Models\label{SUBSEC_GNMs}}

The bridge between preferential attachment and network in-degree distribution is the growing network model. In this subsection we describe a general growing network modelling scheme that includes a number of important growing network models as special cases. Table~\ref{TAB_GNM_Summary} summarises the particular growing network models described in detail below.

We define a \emph{growing network model} as a growing network subject to the following constraint: each edge from among the $m_t$ edges added at time-step $t>1$ connects to a given node of in-degree $k$ from $G_{t-1}$ with probability proportional to the \emph{attachment function} $A(k)$, a time-independent function of $k$ that governs the formation of new connections. In particular, the probability that a said edge connects to some in-degree $k$ node from $G_{t-1}$ is given by
\begin{equation}~\label{EQ_growing_model_transition_rule}
\pi_{t}(k) \propto n_{t-1}(k) \times A(k),
\end{equation}
where $n_{t-1}(k)$ stands for the number of in-degree $k$ nodes in $G_{t-1}$ for $t>1$. It is the form of $A(k)$ together with any structural constraints imposed on the values of $T$, $n_t$, $m_t$, and $m_{t}'$ that defines a growing network model. The attachment rate $\hat{A}(k)$ of a growing network may be rightly regarded as a realisation of an attachment function, $A(k)$, as defined by a compatible growing network model. Note that we allow for multiple edges to occur between nodes in the above formulation of a growing network model as a matter of mathematical convenience. 

\begin{table}[h]
\caption{\textbf{An assortment growing network models relevant to the present work}.}
{\label{TAB_GNM_Summary}}
\centering
\begin{tabular}{llclr}
\textbf{Model Name} & \textbf{Attach. Fn.} & \textbf{Eq.} & \textbf{Deg. Dist.} & \textbf{Ref.} \\ \hline
 Price's model & preferential & (\ref{EQ_Linear_AF}) & power-law & \cite{deSollaPrice1976}  \\
 Jeong's model & preferential & (\ref{EQ_Linear_AF}) & unknown & \cite{Jeong2003}  \\
 Callaway's model & uniform & (\ref{EQ_Uniform_AF}) & exponential & \cite{Na1970, Callaway2001} \\
 Krapivsky's model  & log-linear & (\ref{EQ_Loglinear_AF}) & diverse & \cite{Krapivsky2000, Krapivsky2001} \\
 Redner's model & nonlinear & (\ref{EQ_Nonlinear_AF}) & log-normal & \cite{Redner2005} \\ 
\end{tabular}
\end{table}

In Price's model~\cite{deSollaPrice1976}, or rather, a mild generalisation thereof, the attachment function takes the linear form
\begin{align}\label{EQ_Linear_AF}
A(k) \propto k + 1,
\end{align}
where the unit offset acts as a kind of initial attractiveness, ensuring that zero in-degree nodes stand a fighting chance of acquiring new connections. The form of Eq.~(\ref{EQ_Linear_AF}) makes precise what we mean by a preferential attachment function in this paper. The model definition is completed by taking $n_t = 1$ and assuming the mean value $m$ of the $m_t$'s is constant over time as $t$ becomes large. The average in-degree distribution of networks generated in this manner is known to follow a power-law tail with scaling exponent $\gamma=2+1/m$ in the limit of large~$T$~\cite{Newman2010}. What we will call Jeong's model is the growing network model analog of the bi-epochally resolved growing network construction from the previous subsection. It consists of two networks $G_1$ and $G_2$, i.e., $T=2$.  The preferential attachment function, as defined in Eq.~(\ref{EQ_Linear_AF}), governs the formation of connections between~$m_{2}$ of the $n_{2}$ nodes in $G_2$ with the $n_{1}$ nodes of $G_1$ at time-step $t=2$. Callaway's model, formulated here in the language of Price's model without loss of substance, is the random recursive tree defined by substituting the uniform attachment function
\begin{align}\label{EQ_Uniform_AF}
A(k) \propto 1
\end{align}
into Price's model. Callaway's model has been shown to generate networks with exponentially distributed in-degree distributions~\cite{Na1970, Callaway2001}, and its other properties have been examined at length in the classic literature~\cite{Na1970, Moon1974, Meir1978}. An important Price's model generalisation defined by the log-linear attachment function
\begin{align}\label{EQ_Loglinear_AF}
A(k) \propto (k + 1)^\alpha
\end{align}
for \emph{attachment exponent} $\alpha > 0$ was analysed by Krapivsky, Redner, and Leyvraz~\cite{Krapivsky2000, Krapivsky2001}. They fittingly named their model ``the growing network model,'' but we refer to it as Krapivsky's model in this paper. We have slightly redefined the attachment function from the original $k^\alpha + 1$ for mathematical convenience. Price's model corresponds to the special case when $\alpha=1$. For $0 < \alpha < 1$ the resulting in-degree distribution takes the form of a stretched exponential function~\cite{Krapivsky2001}. For $\alpha > 1$ all nodes connect to a handful of large hubs. Meanwhile the limiting case of $\alpha = 0$ corresponds to Callaway's model. Note that we show the APS citation distribution fitted to the stretched exponential function predicted by Krapivsky's model in Figure~S3.

Finally, Redner writes in passing that the growing network model obtained by substituting the nonlinear attachment function  
\begin{align}\label{EQ_Nonlinear_AF}
A(k) \propto \frac{k+1}{1+\beta \log(k+1)}
\end{align}
with $\beta > 0$ into Price's model generates networks with log-normally distributed in-degree distributions~\cite{Redner2005}. In Supplementary Note~3, we show that Redner's model, as we will call it, generates networks with in-degree distributions that asymptotically follow the log-normal distribution. The proof is adapted from an outline that was kindly supplied to the authors by Redner via email.

\subsection*{The Preferential Attachment Paradox Illustrated on the APS Citation Data\label{SUBSEC_PA_Paradox_Illustrated}}

It is only in virtue of the preceding digression on growing network models that it has at last become possible to cast the preferential attachment paradox in a reasonably technical light. The paradoxical argument runs as follows:
\begin{labeling}{PA Paradox}
\item [Premise 1] A preferential attachment rate gives rise to networks with power-law in-degree distributions. Recall that we have defined a preferential rate of attachment to mean that a growing network model attachment function $A(k)$ increases linearly with $k$.
\item [Premise 2] Measurement suggests a preferential rate of the attachment for the maximally resolved APS citation growing network. In other words, the observed attachment rate $\hat{A}(k)$ is approximately a linear function of $k$. 
\item [Premise 3] The observed APS citation network in-degree distribution is not well-described by a power-law.
\item [Conclusion] That the APS citation network has a power-law in-degree distribution follows from a naive application of Premises~1~and~2.
\item [Paradox] The stated conclusion is in direct contradiction with Premise~3. In fact, measurement suggests that the APS citation network in-degree distribution is better described by a log-normal distribution, than by a power-law.
\end{labeling}
The conclusion is trivially seen to follow from the premises.~Thus it must be the case that some or another premise is either incoherent or outright false. Redner followed this line of reasoning to its contradictory conclusion for the APS citation network. But it is worth noting that the argument applies to any growing network featuring preferential attachment, which culminates in a network with a log-normally distributed in-degree distribution.

\subsection*{The Preferential Attachment Paradox Resolved\label{SEC_PA_Paradox_Resolved}}

We have seen that a network cannot enjoy a log-normal in-degree distribution and have grown in accordance with preferential attachment without apparently contradicting network theory. Yet, we are committed to the view that the APS citation network is endowed with exactly these properties. In this section we will see that a critical examination of the premises underlying the argument leads to a ready explanation of the paradox.

First of all, it may be outright denied that the APS citation distribution is log-normally distributed over its full domain; for to maintain otherwise would blindly disregard the statistical testing outcomes presented in the subsection on modelling the APS citation network citation distribution. This line of objection, while technically correct, does not present an interesting challenge to the argument. The APS citation distribution being well-described by the log-normal in the body of $k$ (i.e. $0 \leq k \leq 150$) turns out to be enough to resolve the paradox. In fact, we will see that the extent to which the log-normal falls short of the APS citation distribution in the extreme tail region of $k$ (i.e. $k \geq 150$) is explained by an equal and opposite departure from an ideal in the APS citation network attachment rate. The upshot is that log-normality assumption may be accepted without prejudice to the argument.

\begin{figure}[!h]
\centering
\includegraphics[width=14.6cm]{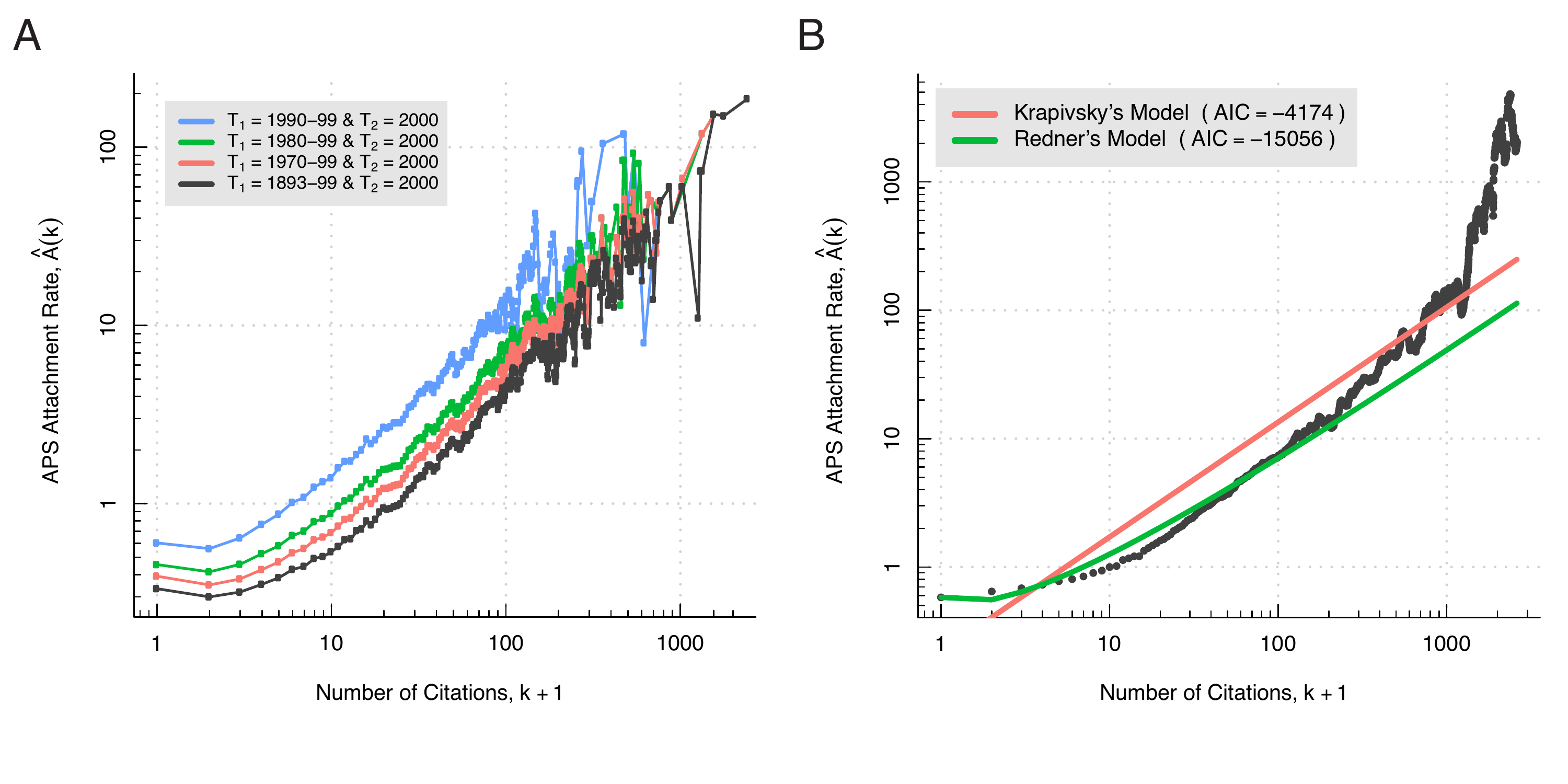}
\caption{\textbf{(A) Nonlinear tendencies in the attachment rates for various bi-epochally resolved APS citation networks} are made apparent on a double logarithmic scale. Four different measurements of $\hat{A}(k)$ may be distinguished by colour in the plot. Each individual measurement is determined by recording how the publications in time interval $T_2=2000$ cite the publications in time intervals $T_1=1990$\thinspace-\thinspace$99$ (blue), $T_1=1980$\thinspace-\thinspace$99$ (green), $T_1=1970$\thinspace-\thinspace$99$ (red), and $T_1=1893$\thinspace-\thinspace$99$ (black), respectively. The data have been averaged over a range of $k \pm 0.025 k$. See Table~S2 and the surrounding text for~details. \textbf{(B) The attachment rate for the maximally resolved APS citation network is best fit by Redner's model}. Shown is the attachment rate for the maximally resolved APS citation network as calculated by Newman's measure (black), the estimated log-linear attachment function of Krapivsky's model (red), and the estimated nonlinear attachment function of Redner's model (green). The best model (the smallest AIC value) is Redner's model. Price's model is not included in the model comparison because it is a special case of Kravipsky's model. The data have been averaged over a range of $k \pm 0.025 k$.\label{FIG_APS_Attachment_Rate_Loglog}}
\end{figure}

The second premise holds that the maximally resolved APS citation network grew in accordance with a preferential rate of attachment. According to Redner's argument, evidence in support of this claim is found in the linear character of the bi-epochally resolved APS citation network attachment rates from Fig.~\ref{FIG_APS_Attachment_Rate}. These results, it will be remembered, were extrapolated to the maximally resolved APS citation network as a whole. Redner's conclusion rests on the assumption that the bi-epochally resolved attachment rates are in fact linear. But in Fig.~\ref{FIG_APS_Attachment_Rate_Loglog}(A) the very same attachment rates are shown plotted on a double logarithmic scale. Visual inspection reveals the log-transformed attachment rates to not strictly adhere to straight line relationships. The linear scale plot of Fig.~\ref{FIG_APS_Attachment_Rate} must therefore conceal the nonlinearities made apparent in the log-log plot, since a straight line must again be such on a double logarithmic scale. This shows how the plotting of attachment rates on a linear scale can be misleading. Thus Redner's extrapolation is thrown into jeopardy, and, as a result, his argument for the truth of the second premise collapses.

The question is whether an explanation of the paradox follows from the manner by which the argument for the second premise fails to apply to the maximally resolved APS citation network. In the remainder of this section we venture to answer the question in the affirmative.

This brings us to the connection between the empirical world of growing networks and the theoretical world of growing network models. In particular, the measuring of a preferential rate of attachment is asserted to be a necessary and sufficient condition for concluding that a growing network is well-modelled by Price's model. This test for Price's model, which at first sight might seem unobjectionable, is revealed to be misleading as soon as efforts are made to formulate it carefully. The trouble stems from implicitly assuming that a preferential rate of attachment is intelligible outside the context of a growing network model. However, preferential attachment is always conditional on a growing network model through not only the laws of edge formation, as defined by the preferential attachment function of Eq.~(\ref{EQ_Linear_AF}), but also the specification of model specific structural constraints. Consequently, the most one can hope to say of a given growing network, even in principle, is that it exhibits preferential attachment with respect to this or that particular growing network model. In other words, a preferential rate of attachment is necessary (but not sufficient) for concluding that Price's model describes a growing network.

In fact, a growing network, $\mathcal{G}$, is obliged to satisfy four conditions in order to comply with Price's model. First, $\mathcal{G}$'s initial network $G_1$ should be small relative to its final network $G$, i.e., $n_1 \ll N$. Practical experience suggests to us that $N \geq 1000 \times \sqrt n_1$ serves as a good rule of thumb, but this by no means rests on a sound theoretical foundation. Second, $\mathcal{G}$ must grow by a single node at each time-step, i.e., $n_t = 1$ for $t>1$. Third, the number of edges $m_t$ added at time-step $t$ must come from a distribution with a fixed mean and finite variance. Fourth, the formation of connections must be governed by the linear attachment function of Eq.~(\ref{EQ_Linear_AF}), i.e., preferential attachment must prevail in the growth of the network. The test for Price's model that is assumed in the second premise ignores all but the last of these conditions.

Let us reinterpret the Fig.~\ref{FIG_APS_Attachment_Rate_Loglog}(A) attachment rates in the light of these new revelations. The first thing to note is that a casual inspection of Table~S2 reveals the corresponding bi-epochally resolved APS citation networks to be in blatant violation of the Price's model structural constraints. They are, however, consistent by definition with the Jeong's model structural constraints. The second thing is that there is a noticeable tendency toward log-linearity in the attachment rates as the $T_2=2000$ articles cite the $T_1=1893$\thinspace-\thinspace$99$ (black), $T_1=1970$\thinspace-\thinspace$99$ (red), $T_1=1980$\thinspace-\thinspace$99$ (green), and $T_1=1990$\thinspace-\thinspace$99$ (blue) articles. In the last case, a log-linear fit is especially not out of the question. It is interesting that we came up with a value of $\hat{\alpha} \approx 0.90$ for the corresponding attachment rate exponent. This value is close to $\hat{\alpha} = 1$, which is the mark of a preferential rate of attachment. So there is even a case to be made for the attachment rate plotted in blue being not only log-linear ($\hat{\alpha} = 0.90$), but also approximately linear ($\hat{\alpha} \approx 1.00$). The same, however, cannot be reasonably maintained of the other attachment rates. The point is that Jeong's measure, as applied by Redner to the APS citation data, may be too crude an instrument to permit for the drawing of subtle distinctions in regard to attachment rate functional form.

All this suggests that it is necessary to take seriously the misspecification of the Price's model structural constraints in order to characterise APS attachment rate functional form. Fortunately, Mark Newman devised a way a to measure attachment rates relative to quite a broad class of growing network models~\cite{Newman2001}. Newman's measure is defined according to
\begin{equation}~\label{EQ_AR_Newman}
\hat{A}(k) \defeq \frac{Z}{W(k)} \sum_{t>1} w_t(k) \frac{m_{t}(k)}{n_{t-1}(k)},
\end{equation}
with weights $w_t(k) = m_{t} \times [n_{t-1}(k) \neq 0 ]$ that have sum $W(k) = \sum_{t > 1} w_t(k)$ ($[P]$ denotes the Iverson bracket for given proposition $P$; $[P]=1$ if $P$ is true and $0$ otherwise) and degree independent normalising constant $Z = \sum_{t > 1} n_{t-1} / m_{t-1}$; the symbol $n_t(k)$ is used to denote the number of in-degree $k$ nodes in $G_t$. Newman's measure is consistent with the Price's model structural constraints, because, in contrast with Jeong's measure, it assumes a time resolution consistent with the model. There are several further points regarding the measure that warrant discussion. First, Newman committed a slight error in his original formulation of the measure, the consequence of which was to introduce a waterfall effect in the large $k$ region of measured attachment rates. The measure defined by Eq.~(\ref{EQ_AR_Newman}) incorporates the correction proposed by Pham~et.~al~\cite{Thong2015} to eliminate this artefact (see Fig.~S4 for a dramatic illustration of the said waterfall effect). Second, it is a pleasant exercise to verify that Eq.~(\ref{EQ_AR_Newman}) reduces to Jeong's measure of Eq.~(\ref{EQ_AR_Jeong}) in the special case of Jeong's model. Third, Newman's measure assumes that the constant of proportionality implicit to Eq.~(\ref{EQ_growing_model_transition_rule}) grows in proportion to the time-step~$t$. This assumption holds true for Price's model which is defined by Eq.~(\ref{EQ_Linear_AF}) with constant $m_t$ on average~\cite{Massen2007, Sheridan2012}. By contrast, the measure is necessarily approximate in the cases of Krapisky's model (unless $\alpha=0$ or~$1$)  and Redner's model.

Figure~\ref{FIG_APS_Attachment_Rate_Loglog}(B) shows what happens when Newman's measure is brought to bear on the maximally resolved APS citation network. The results are striking. Visual inspection makes plain that the nonlinear attachment function from Redner's model provides a better fit to the measured attachment rate, than does the log-linear attachment function from Krapivsky's model. The outcome of a model comparison, in which we used the AIC criteria to select the best model, lends numerical support to this conclusion. The AIC score is $-4174$ for Krapivsky's model and $-15056$ for Redner's model. It follows that Redner's model compares favorably to that of Price, since the latter forms a special case of Krapivsky's model. The resolution to the paradox is now obvious: The APS citation distribution closely follows a log-normal distribution, because the underlying network's growth is closely described by a growing network model (i.e. Redner's model) that predicts just such an outcome. This explains the~paradox.
 
\begin{figure*}[!h]
\centering
\includegraphics[width=146mm]{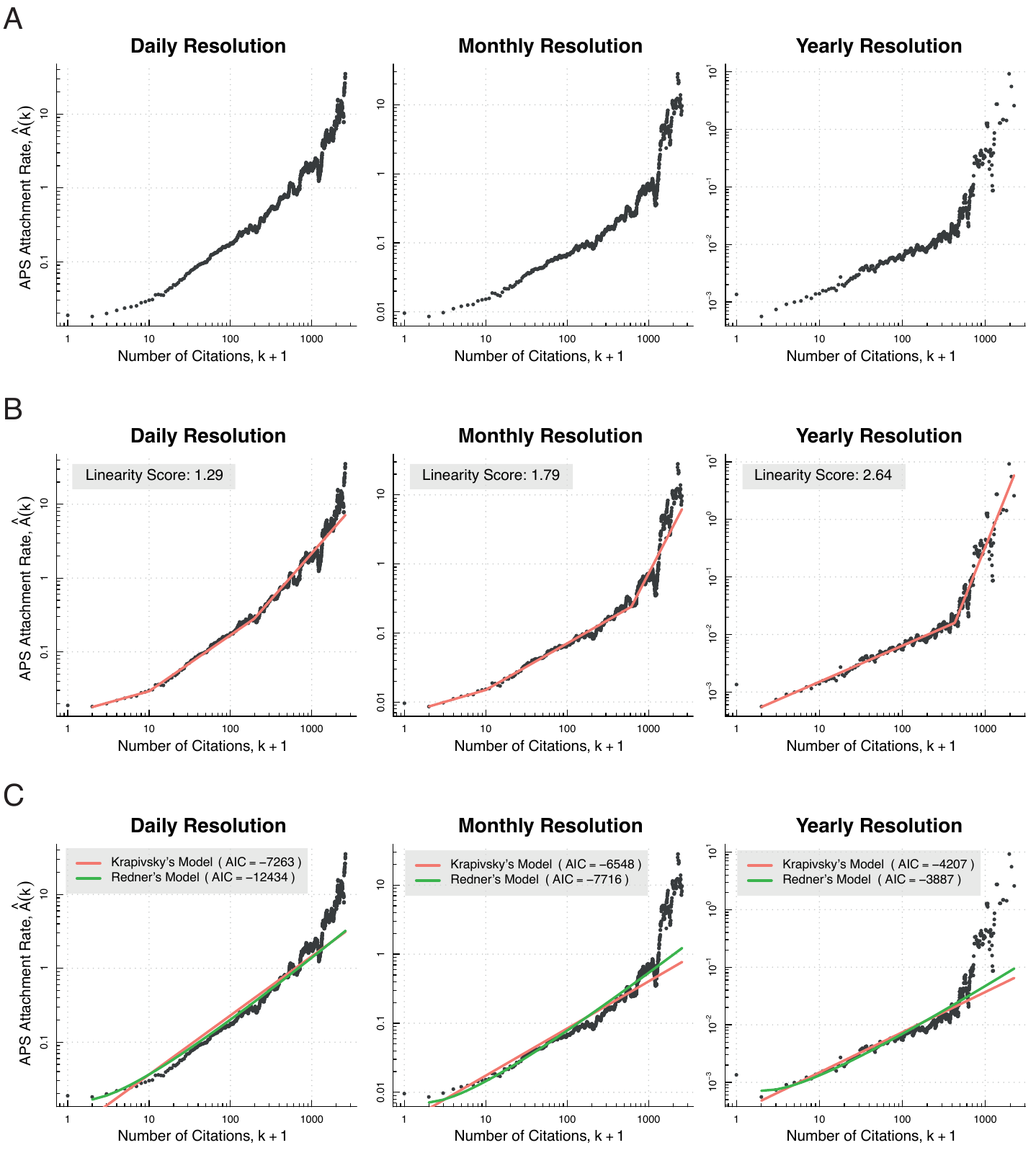}
\caption{\textbf{Overview of the measured attachment rates for growing network representations of the APS journal collection data at daily, monthly, and yearly time resolutions}. (A) The attachment rates plotted in isolation. Each $\hat{A}_k$ was estimated using Newman's measure from APS journal publications dating from July 1893 through 2003 inclusive. The data have been averaged over a range of $k \pm 0.025 k$. (B) The same attachment rates fitted using the segmented linear regression technique discussed in the main text. A larger linearity score reflects a stronger log-linear tendency in a measured attachment rate. The yearly attachment rate is the most log-linear by this score. (C) The attachment rates are fitted by the estimated log-linear attachment function of Krapivsky's model (red), and the estimated nonlinear attachment function of Redner's model (green). The best model (the smallest AIC value) is Redner's model in the case of daily and monthly time resolution, but Krapivsky's model in the case of yearly time resolution. Price's model is not included in the model comparison because it is a special case of Kravipsky's model.\label{FIG_APS_Attachment_Rates_Multiple_Resolution}}
\end{figure*}

It is instructive, as an afterthought, to extend our model comparison to the daily, monthly, and yearly resolved APS citation data. Table~\ref{TAB_Model_Comparison_Summary} shows that the AIC and BIC criteria selects Redner's model over Krapivsky's model in all instances with the single exception of the yearly resolution case. This is interesting because it highlights a tendency toward log-linearity in the APS attachment rate as the time resolution decreases. Fig.~\ref{FIG_APS_Attachment_Rates_Multiple_Resolution} conveys the effect graphically. In Panel~A, the measured attachment rates are plotted for the daily, monthly, and yearly resolved data. Panel~B shows the same attachment rates overlaid with segmented linear regression lines of best fit we calculated using the R package earth~4.4.7~\cite{Milborrow2011}. We defined a log-linearity score heuristic for an attachment rate as the common logarithm of the horizontal component of the longest log-linear segment. Thus the higher the score, the more ``log-linear'' the attachment rate. As expected, attachment rate log-linearity increases with decreasing time resolution, so that the yearly resolved attachment rate is the most log-linear. Panel~C shows the plotted Redner's model and Krapivsky's model attachment functions of best fit. The lesson is that we can expect crudely time resolved data to exhibit a bias toward log-linearity in measured attachment rates. 
 
\begin{table}[!h]
\caption{\textbf{Model comparison results for growing network representations of the APS citation data at various time resolutions}. Shown are AIC and BIC values for the fit of the log-linear attachment function of Krapivsky's model and the nonlinear one of Redner's model to the maximally, daily, monthly, and yearly resolved APS citation data attachment rate, respectively. The best model (the smallest AIC/BIC value) for each level of resolution is indicated in bold. Redner's model best describes the data at the three highest levels of resolution (maximal, daily, and monthly). Krapivsky's model best describes the data at the lowest level of resolution (yearly).}
{\label{TAB_Model_Comparison_Summary}}
\centering
\begin{tabular}{l*{5}{rr}}
\textbf{Resolution} & \textbf{Model} & \textbf{Attach. Fn.} &  \textbf{Eq.} & \textbf{AIC} & \textbf{BIC} \\ \hline
  \multirow{2}{*}{Maximal} & Krapivsky &  Log-linear & (\ref{EQ_Loglinear_AF}) & -4,174 & -4,294\\
  & \textbf{Redner} & \textbf{Nonlinear} & (\ref{EQ_Nonlinear_AF}) & -15,056 & -12,429 \\
 \multirow{2}{*}{Daily} & Krapivsky & Log-linear & (\ref{EQ_Loglinear_AF}) & -7,262 & -7,252\\
  & \textbf{Redner} & \textbf{Nonlinear} & (\ref{EQ_Nonlinear_AF}) & -12,434 & -12,423 \\
 \multirow{2}{*}{Monthly} & Krapivsky & Log-linear & (\ref{EQ_Loglinear_AF}) & -6,548 & -6,538 \\
  & \textbf{Redner} & \textbf{Nonlinear} & (\ref{EQ_Nonlinear_AF}) & -7,716 & -7,706\\
 \multirow{2}{*}{Yearly} & \textbf{Krapivsky} & \textbf{Log-linear} & (\ref{EQ_Loglinear_AF}) & -4,207 & -4,198 \\
  & Redner & Nonlinear & (\ref{EQ_Nonlinear_AF}) & -3,887 & -3,878 \\  
\end{tabular}
\end{table}
 
All that remains is to tie up a few loose ends. First, we have asserted that confining ourselves to the range $0 \leq k \leq 150$ would be sufficient to explain the paradox. For justification, observe how the maximally resolved APS citation network attachment rate plotted in Fig.~\ref{FIG_APS_Attachment_Rate_Loglog}(B) overshoots the Redner's model attachment function after about $k \geq 150$, and the APS citation distribution plotted in Fig.~\ref{FIG_APS_Citation_Distribution} overshoots the log-normal distribution after about $k \geq 150$. These effects are two sides of the same coin: the attachment rate overshooting the predicted attachment function (i.e. $k \geq 150$ nodes acquiring citations at a higher expected rates) automatically leads to the citation distribution overshooting in the log-normal distribution (i.e. $k \geq 150$ nodes are more highly connected than predicted by the log-normal). Thus the lack of agreement between theory and observation can be understood within the modelling framework we have presented, and does not detract from our arguments. Second, the maximally resolved APS citation data, on which we rely to explain the paradox, is consistent with the constant $m_t$ on average assumption on which the models we have considered here depend (i.e. $m_t = 1$ for all $t$). However, it is important to point out that this assumption is violated for more coarse time resolutions. For example, the number of APS articles have grown exponentially over time with a doubling rate of about $6.5$ years. And lastly, the matter of whether the attachment rate remains constant over time in the case of the APS citation data merits some consideration, since this is an assumption of our models. To test the assumption, we partitioned the data from $1901$ to $2000$ into four non-overlapping time windows (i.e. $1901$\thinspace-\thinspace$74$, $1974$\thinspace-\thinspace$88$, $1988$\thinspace-\thinspace$95$, $1995$\thinspace-\thinspace$2000$) and estimated the attachment exponent separately in each case using Newman's method at maximal time resolution. The time windows were selected such that the number of articles are equally distributed. The corresponding estimates for $\alpha$ (i.e. $0.97$, $0.94$, $1.05$, \& $1.06$, respectively) lend credence to the notion that the constant attachment rate assumption holds at least approximately true.

\section*{Discussion\label{SEC_Discussion}}

The main purpose of this paper has been to resolve the preferential attachment paradox. Our proposed resolution highlights various pitfalls that the working network scientist would do well to avoid when measuring preferential attachment.

First, we have called attention to the basic fact that an attachment rate is always measured relative to this or that growing model. Granted, this observation is not particularly important as regards the brute assessment of whether or not real-world network attachment rates increase on average with node degree. Recall that this is one way to define preferential attachment. The measurement procedures that Mark Newman~\cite{Newman2001} and Jeong et. al~\cite{Jeong2003} proposed in the early 2000s have proved adequate for confirming this form of preferential attachment for numerous instances as summarised in other sources~\cite{Sheridan2012, Thong2015}. The same holds true of more recent measurement procedures~\cite{Massen2007, Gomez2011, Kunegis2013, Thong2015}. But the situation is completely different for the characterisation of attachment rate functional form. In the present work we have seen that the APS citation network attachment rate is better modelled by a nonlinear function under maximal time resolution, and a log-linear function under yearly resolution. This serves as a cautionary tale when it comes to making model-free statements about attachment rate functional form. Second, we have taken pains to state the importance of using the corrected version of Newman's method~\cite{Thong2015} when assessing attachment rate function form at fine time resolutions. Third, the importance of plotting attachment rates on a double logarithmic scale cannot be overstated in light of the striking contract between the plots of Figs.~\ref{FIG_APS_Attachment_Rate} and ~\ref{FIG_APS_Attachment_Rate_Loglog}(A).

On a different note, we would be remiss not to comment on the conspicuous lack of statistical formalism employed in the analysis of attachment rate data. The contrast in technical sophistication between the manners in which degree distributions and attachment rates are characterised in the literature is striking. Analysing the APS citation distribution was straightforward thanks to the statistical formalism of Clauset et al.~\cite{Clauset2009} as implemented in the poweRlaw R package~\cite{Gillespie2015}. More generally, the standardisation of fitting power-laws and other heavy-tailed forms to observed degree distributions was a direct outcome of Clauset et al.~\cite{Clauset2009}. No comparable formalism exists for attachment rate analysis to our knowledge. Although important strides in the modelling of citation dynamics are found in the work of Eom and Fortunato~\cite{Eom2011}, and Golosovsky and Solomon~\cite{Golosovsky2012, Golosovsky2017}. This is an intolerable state of affairs seeing that attachment rate and degree distribution are a package deal in so far as growing networks are concerned. An easy-to-use statistical toolkit is needed for fitting and comparing established growing network model attachment functions to observed attachment rates. Fortunately, it should be possible to adapt the maximum likelihood estimation methods and goodness-of-fit tests described in Clauset et al.~\cite{Clauset2009} to this purpose. Implementing the proposed methodology in Python and R would go a long way to streamline the analysis of attachment rate data in academic publications.

Lastly, there is a pressing need for a review paper on the measurement of the chief processes describing how complex networks change over time. The measuring of preferential attachment in growing networks, which has so preoccupied our thinking in the present work, is part of a larger enterprise to measure nothing short of all conjectured network evolutionary processes. Preferential attachment is one of many such processes to have been conjectured, including node fitness~\cite{Bianconi2001}, node duplication coupled with edge rewiring~\cite{Pastor-Satorras2003}, homophily~\cite{McPherson2001}, topological distance~\cite{Newman2001}, and node birth/death processes~\cite{Dorogovtsev2000}. At least three good reviews have been written on generative network models~\cite{Albert2002, Boccaletti2006, Holme2015}, but none on the subject of measuring the processes they embody in real-world networks. It is high time for a survey of the methodological landscape and critical exposition of real-world findings in this area be undertaken.

\bibliography{references}

\section*{Acknowledgements}

We kindly thank Sidney Redner for supplying us with a sketch of the proof which appears in Supplementary Note 3 and Thong Pham for some helpful discussions.

\section*{Author contributions statement}

P.S. and T.O. conceived the analysis,  P.S. conducted the analysis. P.S. and T.O. wrote the manuscript. 

\section*{Competing interests}

The authors declare that they have no competing interests.

\end{document}